\documentclass{article}
\usepackage{ijcai26}
\usepackage{natbib}
\usepackage{times}
\usepackage{soul}
\usepackage{url}
\usepackage[hidelinks]{hyperref}
\usepackage[utf8]{inputenc}
\usepackage[small]{caption}
\usepackage{graphicx}
\usepackage{amsmath}
\usepackage{amsthm}
\usepackage{booktabs}
\usepackage{algorithm}
\usepackage{algorithmic}
\usepackage[switch]{lineno}
\usepackage{xcolor}
\usepackage{booktabs}
\usepackage{amsmath,amssymb}
\usepackage{multirow}
\usepackage{threeparttable}
\usepackage{tcolorbox}
\tcbset{
    colback=gray!5,
    colframe=black!60,
    boxrule=0.5pt,
    arc=3pt,
    left=6pt,
    right=6pt,
    top=6pt,
    bottom=6pt,
}

\title{Cross-Stock Predictability via LLM-Augmented Semantic Networks}

\author{
Yikuan Huang$^{1,2,*}$
\and
Zheqi Fan$^{1,2,*}$
\and
Kaiqi Hu$^{3}$
\and
Yifan Ye$^{4}$
\\
\affiliations
$^1$Division of EMIA,
Hong Kong University of Science and Technology, Hong Kong SAR\\
$^2$Thrust of FinTech,
Hong Kong University of Science and Technology, Guangzhou, China\\
$^3$Rutgers Business School,
Rutgers University–New Brunswick, New Jersey, U.S.A.\\
$^4$Faculty of Business and Management, Beijing Normal–Hong Kong Baptist University, Zhuhai, China\\
\textsuperscript{*}Corresponding authors.
\emails
\{yk.huang, zheqi.fan\}@connect.ust.hk
\and
kh962@scarletmail.rutgers.edu
\and
yifanye@bnbu.edu.cn
}

\begin{document}

\maketitle

\begin{abstract}
    Text-based financial networks are increasingly used to study cross-stock return predictability. A common approach constructs links from similarities in firms' disclosure embeddings, but such networks often contain spurious edges because textual proximity does not necessarily imply economic connection. We propose a two-stage framework that first builds a sparse candidate graph from 10-K embeddings and then uses a large language model to classify and filter candidate edges according to their economic relations. The refined graph is used to aggregate pair-level mean-reversion signals into stock-level trading signals with relation-aware and distance-based weights. In a backtest on S\&P~500 constituents from 2011 to 2019, LLM-based edge filtering improves the long-short Sharpe ratio from 0.742 to 0.820 and reduces maximum drawdown from $-$10.47\% to $-$7.85\%. These results suggest that LLM-based reasoning can improve the economic fidelity of text-derived financial networks and strengthen cross-stock predictability.

\end{abstract}

\section{Introduction}
\label{sec:introduction}

Cross-stock predictability---the phenomenon that a stock's future return can be partially predicted by the contemporaneous behaviour of economically linked firms---is one of the most robust empirical regularities in asset pricing~\citep{cohen2008economic, menzly2010market}. This predictability arises because information diffuses gradually across interconnected firms: a demand shock to a downstream manufacturer, for instance, eventually propagates to its upstream suppliers, but often with a delay that creates exploitable price dislocations. Capitalising on this phenomenon requires a reliable map of economic linkages---a \textit{network} that captures who is genuinely connected to whom.

Recent advances in natural language processing have made it possible to construct such networks at scale from corporate textual disclosures. By encoding 10-K filings into dense embedding vectors and linking each firm to its top-$K$ most similar peers, one obtains a semantic graph that updates annually and covers the full public equity universe. However, this approach introduces a subtle but critical problem: \textbf{spurious correlations}. Embedding models capture \textit{topical similarity} rather than \textit{economic causality}---two firms may discuss identical macroeconomic headwinds (e.g., ``supply chain disruptions,'' ``regulatory uncertainty'') without sharing any direct business interaction. Aggregating cross-stock signals over such noisy edges dilutes the genuine predictive content and degrades portfolio performance. Consider two firms in unrelated industries that both devoted large sections of their filings to pandemic-related risks: their embeddings would be close, yet a price divergence between them carries no mean-reversion information.

In this paper, we propose a \textit{Retrieve-then-Reason} framework that addresses the spurious correlation problem by introducing a large language model (LLM) as an economic reasoner. The first stage performs a lightweight embedding similarity search to generate a candidate graph with $O(NK)$ edges, making the subsequent reasoning step computationally tractable. The second stage queries an LLM to classify each candidate edge into one of six mutually exclusive relationship categories (e.g., \textit{competitor}, \textit{supply chain}, \textit{peer}). Competitor edges---where price divergence more likely reflects structural market-share shifts---are removed, while substitute edges are down-weighted. The refined graph is then used to construct stock-level signals via a novel per-stock softmax aggregation weighted by Gatev distance, so that pairs exhibiting tighter historical co-movement receive greater influence.

We evaluate the framework on U.S. equities (S\&P~500 universe, 2011--2019) and demonstrate three main findings: (i) adding LLM filtering raises the long-short Sharpe ratio from 0.742 to 0.820 while reducing maximum drawdown by 262 basis points; (ii) the semantic network substantially outperforms both random graphs and SIC-based industry networks; and (iii) Fama--French factor regressions confirm a statistically significant daily alpha not subsumed by conventional risk premia.

\paragraph{Contributions.} Our main contributions are threefold:
\begin{enumerate}
    \item We propose a novel \textit{Retrieve-then-Reason} framework that leverages LLMs to filter spurious correlations in embedding-based financial networks, producing a high-fidelity economic graph with semantically labelled edges.
    \item We design a relation-aware, distance-weighted signal aggregation mechanism that maps pair-level z-scores to stock-level cross-sectional signals, with weights informed by both economic relationship types and historical price co-movement.
    \item We conduct an extensive empirical evaluation including component ablation, parameter sensitivity analysis, and risk-factor regression, demonstrating statistically significant portfolio improvements attributable to the LLM reasoning module.
\end{enumerate}

\section{Related work}
\label{sec:literature}

Our work sits at the intersection of empirical asset pricing and natural language processing, specifically bridging cross-stock predictability with LLM-based reasoning.

\subsection{Cross-Stock Predictability and Economic Linkages}
A rich literature in finance documents that the returns of economically linked firms can predict the future returns of a focal firm, a phenomenon often attributed to investor inattention and slow information diffusion. Researchers have identified various predefined linkages, including supply chains \citep{cohen2008economic}, industry classifications \citep{moskowitz1999do}, geographic proximity \citep{parsons2020geographic}, and shared analyst coverage \citep{ali2020shared}.
Recent evidence further documents cross-stock predictability based on return similarity \citep{chen2026cross}.

Recent advancements in this field emphasize the critical need to dissect the nature of these linkages. \cite{yan2023cross} theoretically characterize cross-stock linkages into symmetric components (driven by common factor momentum) and asymmetric components (unidirectional lead-lag relationships). Building on this, \cite{li2025how} demonstrate that the symmetric component often leads to long-term return reversals (reflecting overreaction), whereas the asymmetric component---such as liquid stocks leading illiquid ones---genuinely reflects investor underreaction. Furthermore, \cite{avramov2025dual} introduce the concept of ``Dual Peer Effects,'' showing that both the collective strength of a peer group and a firm's relative position within it contribute to predictability. While these studies highlight the importance of high-fidelity, directional economic networks, they primarily rely on static or manually curated classifications (e.g., SIC codes or Compustat customer-supplier segments), which often lag behind dynamic market realities.

\subsection{Data-Driven Peer Networks and Their Limitations}

Recent studies construct firm networks from high-dimensional textual and patent data rather than relying on predefined economic classifications. 
\citet{hoberg2018text} develop text-based industry momentum using 10-K product descriptions, and \citet{eisdorfer2022competition} extract competitor networks from regulatory filings. 
Patent-based approaches measure innovation similarity or technological proximity to predict returns \citep{bekkerman2023effect,lee2019technological}, while \citet{lee2024production} document momentum spillovers driven by production complementarity.
Alongside text and patent data, recent methodologies also employ graph learning to infer dynamic momentum spillover networks directly from pricing data across diverse asset classes \citep{pu2023network}.

These approaches demonstrate the value of unstructured data, but they primarily rely on similarity metrics—such as cosine similarity in embedding space or overlap in patent classifications—to infer economic connections. 
Such representations reduce inter-firm relationships to continuous proximity scores and implicitly assume that greater similarity corresponds to stronger economic linkage. 
However, textual or technological similarity does not uniquely identify the nature of the underlying relationship. 
Firms with similar disclosures may be direct competitors (substitutes), vertically related partners, or strategic complements. 
Pooling these heterogeneous links into a single similarity-based network may therefore introduce structural noise and attenuate return predictability.

We depart from distance-based inference and introduce an LLM-driven relational framework. 
Rather than measuring proximity in embedding space, we leverage a large language model to perform contextual semantic reasoning over 10-K disclosures and explicitly classify inter-firm references into structured economic categories (e.g., \textit{Competitor}, \textit{Supply Chain}, \textit{Complementary}). 
This design shifts the network construction problem from unsupervised similarity estimation to supervised semantic categorization. 
By transforming unstructured financial text into a typed relational graph, our framework disentangles economically distinct link types and enables selective filtering when constructing trading signals, thereby improving interpretability and reducing spurious correlations inherent in similarity-based networks.

\subsection{LLMs for Financial Reasoning}
Large Language Models (LLMs) have shown remarkable capabilities in financial text analysis, moving beyond simple sentiment analysis to complex reasoning tasks and interpretable internal representations \citep{wu2023bloomberggpt, kong2024large1,kong2024large2, chen2026financial,xie2024finben}. Recent applications include using LLMs for direct stock return prediction \citep{chen2023expected, lopezlira2023can}, macroeconomic forecasting \citep{chen2025chatgpt}, corporate event analysis \citep{xie2023pixiu}, extracting structured investor beliefs \citep{gao2024structured}, and tracking evolving semantic signals in corporate disclosures \citep{choi2025text}.
Furthermore, the frontier of LLM application is shifting from passive analysis to agentic AI frameworks that autonomously discover and validate systematic trading factors \citep{huang2026beyond}.
However, recent studies raise critical concerns regarding direct economic forecasting, warning that the apparent predictive success of LLMs may largely stem from data contamination and memorization rather than genuine foresight \citep{lopezlira2025memorization}. 

Our work extends this paradigm and circumvents the memorization problem by deploying LLMs not as direct return predictors, but as \textbf{semantic reasoning agents for network refinement}. Unlike traditional NLP approaches that rely on unweighted embedding similarities, our \textit{Retrieve-then-Reason} framework aligns with the financial literature's call for distinguishing the \textit{nature} of cross-stock relationships \citep{yan2023cross}. By prompting the LLM to explicitly classify edges into categories (e.g., Competitor, Supply Chain, Complementary) and filtering out noisy competitor links, we construct a high-fidelity economic graph. This approach directly addresses the limitations of embedding-based networks and mitigates the lookahead bias associated with direct LLM forecasting, providing a structurally sound foundation for cross-stock signal aggregation.

\section{Problem Formulation}
\label{sec:problem}

\subsection{Notation}

Let $\mathcal{V} = \{v_1, \dots, v_N\}$ denote a universe of $N$ stocks. For each stock $v_i$, we observe daily returns $r_{i,t}$ over a time horizon $[1, T]$. We define the \textit{normalized price} of stock $v_i$ starting from a reference date $t_0$ as the cumulative product of gross returns:
\begin{equation}
    P_{i,t} = \prod_{\tau=t_0}^{t} (1 + r_{i,\tau}).
\end{equation}

Each stock is further associated with a textual embedding $\mathbf{h}_i \in \mathbb{R}^D$, derived from its corporate disclosure documents (e.g., 10-K filings). These embeddings serve as a compressed representation of the firm's business profile.

\subsection{Network-Based Cross-Stock Predictability}

Cross-stock predictability exploits the fact that information diffuses gradually across economically linked firms~\citep{cohen2008economic, menzly2010market}. The key premise is that the future return of a target stock $v_i$ can be partially predicted by observing the current states of its economic peers. Formally, we represent these linkages as a graph $\mathcal{G} = (\mathcal{V}, \mathcal{E})$, and the cross-stock predictive signal for stock $v_i$ is constructed by aggregating information from its neighbourhood:
\begin{equation}
    S_{i,t} = \text{Agg}\bigl(\{(v_j, r_j, P_j) \mid e_{i,j} \in \mathcal{E}\}\bigr).
    \label{eq:signal_abstract}
\end{equation}

The quality of this signal critically depends on the fidelity of $\mathcal{E}$: an ideal edge set captures only genuine economic interactions, so that price dislocations among connected stocks reflect temporary inefficiencies rather than permanent divergences.

\subsection{The Spurious Correlation Problem}

A natural approach to constructing $\mathcal{E}$ is to exploit textual similarity. For each stock pair, the cosine similarity between their embeddings is computed:
\begin{equation}
    \mathrm{sim}(v_i, v_j) = \frac{\mathbf{h}_i^{\!\mathsf{T}} \mathbf{h}_j}{\|\mathbf{h}_i\| \cdot \|\mathbf{h}_j\|},
\end{equation}
and a top-$K$ neighbour graph $\mathcal{G}_{emb} = (\mathcal{V}, \mathcal{E}_{emb})$ is formed by retaining the $K$ most similar peers for each node. While efficient and scalable, this approach suffers from a fundamental limitation: textual embeddings capture \textit{topical similarity} rather than \textit{economic causality}. Two firms may discuss similar macroeconomic risks (e.g., ``supply chain disruptions,'' ``regulatory uncertainty'') in their filings without having any direct business interaction. We refer to these noisy edges as \textbf{spurious correlations}. Aggregating signals over such edges dilutes the genuine predictive content and degrades portfolio performance.

\subsection{Objective}

The central objective of this work is to refine the candidate graph $\mathcal{G}_{emb}$ into a high-fidelity economic network $\mathcal{G}_{ref} = (\mathcal{V}, \mathcal{E}_{ref})$, where $\mathcal{E}_{ref} \subseteq \mathcal{E}_{emb}$, such that the retained edges correspond to substantive economic relationships. We further seek to assign each retained edge a semantic label that characterises the nature of the relationship, enabling relation-aware signal construction. The core research question is:

\begin{quote}
    \textit{Can a large language model, by reasoning over corporate textual information, effectively distinguish genuine economic linkages from spurious ones, thereby improving cross-stock predictability?}
\end{quote}

\section{Proposed Framework}
\label{sec:framework}

\begin{figure*}[t]
    \centering
    \includegraphics[width=\textwidth]{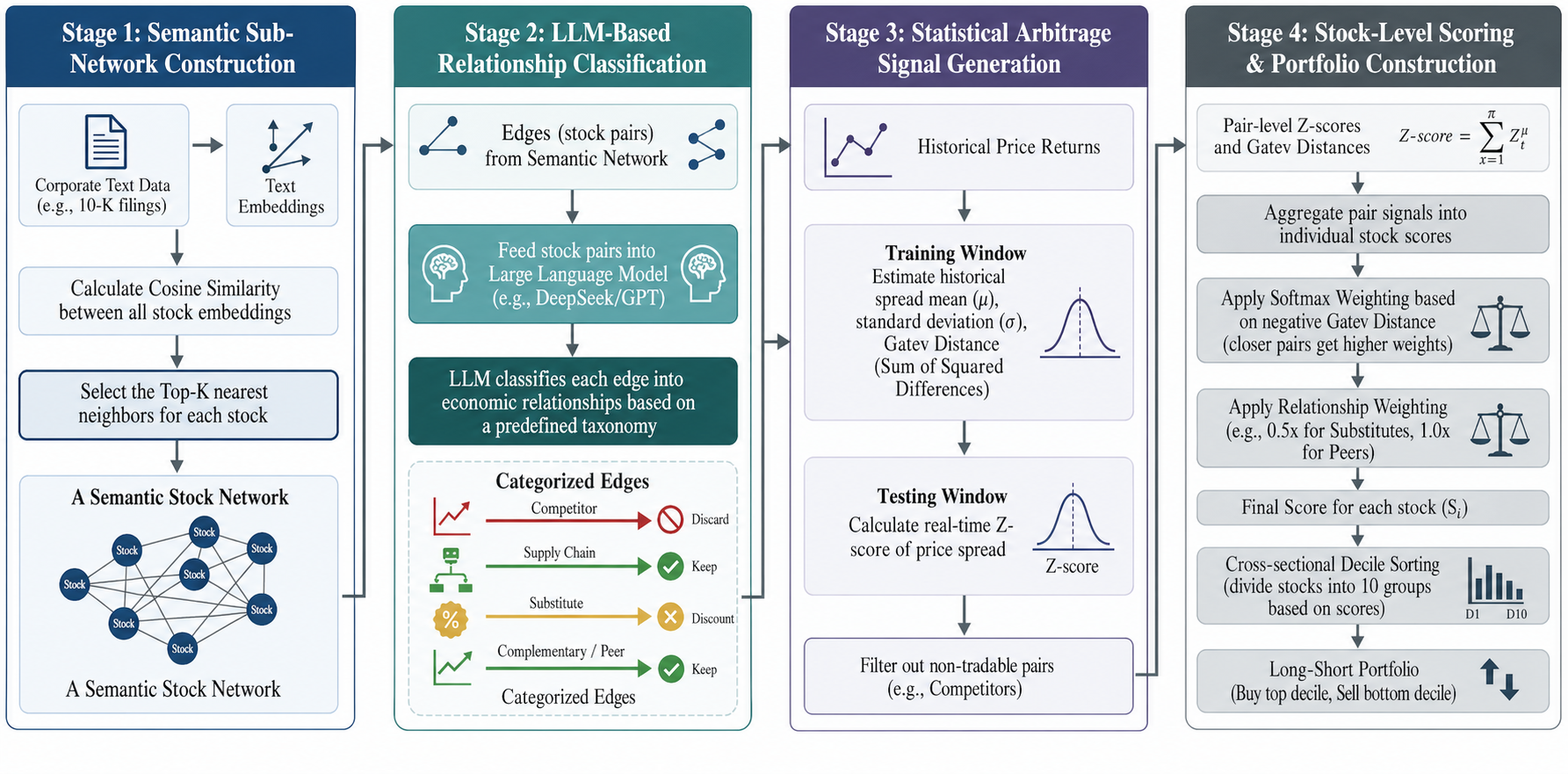}
    \caption{Overview of the proposed framework.}
    \label{fig:framework}
\end{figure*}

We propose a two-stage framework---\textit{Retrieve then Reason}---that first efficiently identifies candidate stock pairs via embedding similarity, then deploys a large language model to verify and classify their economic relationships. The refined network is subsequently used to construct cross-stock predictive signals. Figure~\ref{fig:framework} illustrates the overall architecture.

\subsection{Stage 1: Candidate Graph Generation}
\label{sec:stage1}

Given the embedding matrix $\mathbf{H} = [\mathbf{h}_1, \dots, \mathbf{h}_N]^{\!\mathsf{T}} \in \mathbb{R}^{N \times D}$, we compute the full pairwise cosine similarity matrix and, for each stock $v_i$, retain edges to its top-$K$ most similar peers. The resulting candidate graph $\mathcal{G}_{emb} = (\mathcal{V}, \mathcal{E}_{emb})$ is made undirected by symmetrisation: an edge $(v_i, v_j)$ is included if either $v_j$ ranks in $v_i$'s top-$K$ or vice versa.

This stage is lightweight and scalable, reducing the number of edges from $O(N^2)$ to $O(NK)$, which makes the subsequent LLM reasoning computationally tractable.
\subsection{Stage 2: LLM-Augmented Edge Classification}

Given the candidate edge set $E_{\text{emb}}$, we employ a large language model (LLM) to classify the economic relationship between each pair of firms. Unlike prior approaches that rely solely on embedding similarity, this stage performs structured semantic reasoning over corporate disclosures to assign economically meaningful labels.

\paragraph{Input Construction.}
For each candidate edge $(v_i, v_j)$, we construct a structured prompt using textual information extracted from firms’ annual 10-K filings (fiscal year $y-1$). Specifically, we include three components for each firm:
(i) a truncated business description from Item 1,
(ii) product and segment-related disclosures, and
(iii) competitor-related sentences identified via regular expressions (e.g., phrases surrounding ``compete with'').

To reduce input length and ensure comparability across firms, each section is truncated to a fixed token budget. Importantly, firm identities are anonymized in the prompt (e.g., ``Firm A'' and ``Firm B''), preventing the model from relying on memorized knowledge of specific tickers and encouraging reasoning based solely on disclosure content.

\paragraph{Prompt Design.}
The LLM is instructed to act as an industry analyst and classify the relationship between two firms based only on the provided disclosures. The model must select exactly one label from a predefined taxonomy:
\{\texttt{competitor}, \texttt{supply\_chain}, \texttt{complementary}, \texttt{substitute}, \texttt{peer}, \texttt{unrelated}\}.
The output is required in structured JSON format, including both the predicted label and supporting evidence spans from each firm's disclosure. This design enforces deterministic, interpretable outputs and facilitates downstream auditing.

\paragraph{Relation Taxonomy.}
The classification schema is designed to reflect economically distinct channels of cross-stock interaction. In particular, it distinguishes between relationships that are likely to generate mean-reversion signals (e.g., supply chain, complementary) and those that reflect structural competition (e.g., competitors). This distinction is critical for filtering edges in subsequent stages.

\paragraph{Implementation Details.}
We use DeepSeek-Chat via API with temperature set to zero to ensure deterministic outputs. Queries are batched to improve efficiency, and results are cached across rolling windows to avoid redundant inference. The total number of queries scales as $O(NK)$, making the approach computationally tractable given the sparsity of the candidate graph.

\paragraph{Discussion.}
This stage transforms the network construction problem from unsupervised similarity estimation into a structured semantic classification task. By leveraging LLM-based reasoning over firm disclosures—rather than relying on embedding proximity alone—we aim to reduce spurious edges and improve the economic fidelity of the resulting network.

\subsection{Cross-Stock Signal Construction}
\label{sec:signal}

Given the refined graph $\mathcal{G}_{ref}$, we construct a predictive signal for each stock using a rolling-window procedure.

\paragraph{Training phase.}
For each edge $(v_i, v_j) \in \mathcal{E}_{ref}$, we compute the normalised price spread during the training window $[t_0, t_1]$:
\begin{equation}
    s_{ij,t} = P_{i,t} - P_{j,t}, \quad t \in [t_0, t_1]
\end{equation}
and estimate the spread's historical mean $\mu_{ij} = \overline{s}_{ij}$ and standard deviation $\sigma_{ij}$. We also record \cite{gatev2006pairs}'s distance 
$d_{ij} = \sum_{t=t_0}^{t_1} s_{ij,t}^2,$ 
which quantifies the overall closeness of the two normalised price paths during training.

\paragraph{Test phase.}
During the test window $[t_1{+}1, t_2]$, we compute the z-score for each pair:
\begin{equation}
    z_{ij,t} = \frac{s_{ij,t} - \mu_{ij}}{\sigma_{ij}}.
\end{equation}

A large positive $z_{ij,t}$ indicates that $v_i$ has become relatively expensive compared to $v_j$, predicting a future mean reversion where $v_i$ underperforms and $v_j$ outperforms.

\paragraph{Per-stock signal aggregation.}
Each stock participates in multiple edges. We aggregate pair-level z-scores into a stock-level signal $S_{i,t}$ using the sign convention:
\begin{equation}
    S_{i,t} = \sum_{(i,j) \in \mathcal{E}_{ref}} \bigl(-z_{ij,t} \cdot w_{ij}^{(i)}\bigr), \quad
    S_{j,t} = \sum_{(i,j) \in \mathcal{E}_{ref}} \bigl(+z_{ij,t} \cdot w_{ij}^{(j)}\bigr).
    \label{eq:agg}
\end{equation}

When the spread is abnormally high ($z > 0$), stock $v_i$ receives a negative (sell) signal and stock $v_j$ receives a positive (buy) signal, reflecting the mean-reversion hypothesis.

The weight $w_{ij}^{(i)}$ is computed via a per-stock softmax over negative distances used by \cite{gatev2006pairs}, scaled by the relation weight:
\begin{equation}
    w_{ij}^{(i)} = \omega_r \cdot n_i \cdot \frac{\exp(-d_{ij})}{\displaystyle\sum_{(i,k) \in \mathcal{E}_{ref}} \exp(-d_{ik})}
    \label{eq:weight}
\end{equation}
where $n_i = |\{(i,k) \in \mathcal{E}_{ref}\}|$ is the degree of stock $v_i$ in the refined graph. The softmax assigns higher weight to pairs with smaller historical distances (tighter co-movement), while the $n_i$ scaling factor preserves the aggregate signal magnitude across stocks with varying node degrees.

\subsection{Portfolio Construction}
\label{sec:portfolio}

At each rebalancing date $t$, we sort all stocks cross-sectionally by their signal $S_{i,t}$ and partition them into $G$ equal-sized groups. Group~1 contains stocks with the lowest scores (predicted to decline) and Group~$G$ contains those with the highest scores (predicted to appreciate). The long-short portfolio goes long Group~$G$ and short Group~1, and its return $r_t^{LS} = \bar{r}_t^{G} - \bar{r}_t^{1}$ serves as the primary measure of cross-stock predictability. We report annualised return, annualised volatility, Sharpe ratio, and portfolio turnover.

\section{Experiments}
\label{sec:experiments}

\subsection{Data and Setup}

\paragraph{Stock universe.}
We use daily returns of U.S. common stocks from the Center for Research in Security Prices (CRSP) database. The full return panel spans July~2007 to December~2020, covering 14{,}218 securities and 3{,}412 trading days. We restrict the investable universe to firms appearing in the S\&P~500 constituent list, yielding approximately 605 unique PERMNOs.

\paragraph{Textual embeddings.}
For each calendar year $y$, we obtain 768-dimensional embedding vectors derived from firms' annual risk-factor disclosures (10-K filings), encoded using a pre-trained language model. After matching embeddings to the CRSP universe, approximately 497 stocks per year have valid embeddings.

\paragraph{Rolling-window backtest.}
We evaluate out-of-sample performance over the period January~2011 to December~2019 using a rolling-window design. Each window consists of a 180-trading-day training period for estimating spread parameters, followed by a 2-month test period for generating signals and measuring portfolio returns. Embeddings from year $y{-}1$ are used for the windows whose training period starts in year $y$, ensuring no look-ahead bias in graph construction.

\paragraph{Portfolio construction.}
At each rebalancing date, stocks are sorted cross-sectionally by their aggregated signal $S_{i,t}$ into $G = 5$ equal-sized quintile groups. We form a long-short portfolio that goes long Group~5 (highest predicted return) and shorts Group~1 (lowest predicted return). Portfolios are equal-weighted within each group and rebalanced daily unless otherwise noted.

\paragraph{LLM configuration.}
For the LLM-augmented edge classification (Stage~2), we use DeepSeek-Chat via API with temperature set to 0 for deterministic outputs. Results are cached to ensure reproducibility across rolling windows.

\paragraph{Evaluation metrics.}
We report annualised return ($r^{ann}$), annualised volatility ($\sigma^{ann}$), Sharpe ratio ($SR$), maximum drawdown ($MDD$), annualised turnover ($TO^{ann}$), and the Newey--West $t$-statistic ($t_{NW}$) for the long-short portfolio.

\subsection{Main Results}

\begin{figure*}[t]
    \centering
    \includegraphics[width=\textwidth]{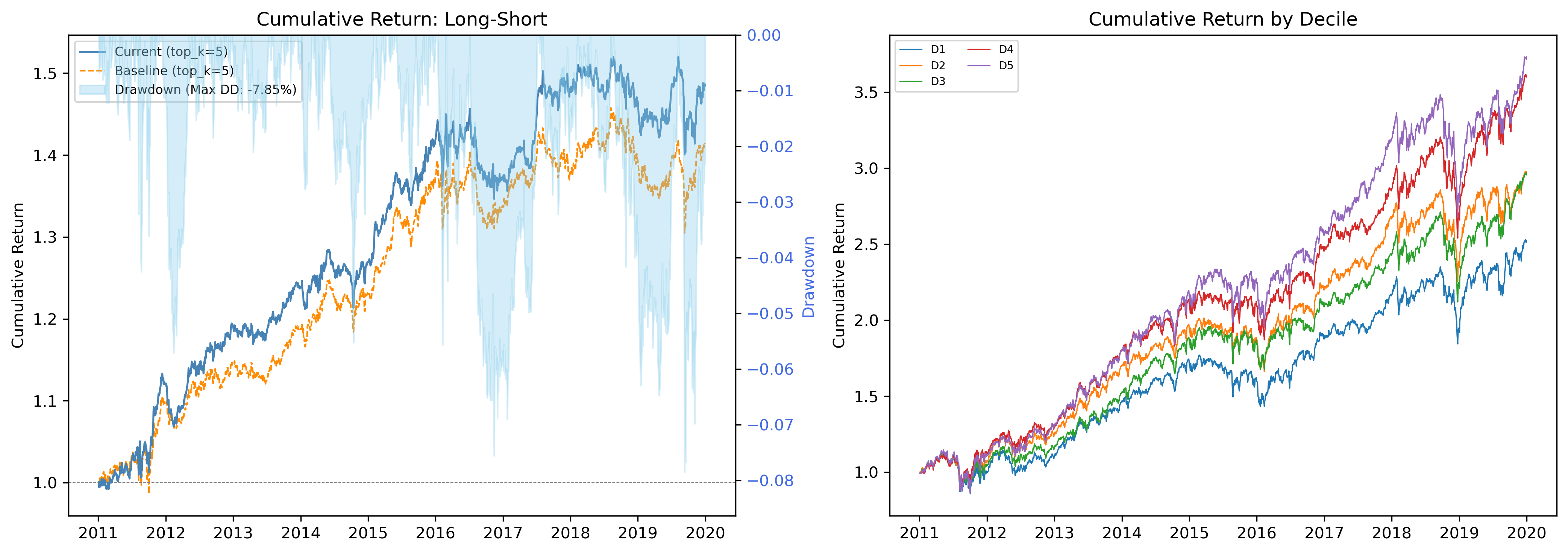}
\caption{Cumulative returns of long-short portfolio and quintile sorting portfolios.}
    \label{fig:cumret}
\end{figure*}

Table~\ref{tab:main_results} reports the long-short portfolio performance under five configurations, designed to isolate the contribution of each component.

\begin{table*}[t]
\centering
\caption{Core Component Ablation: Long-Short Portfolio Performance}
\label{tab:main_results}
\footnotesize
\begin{tabular}{lcccccc}
\toprule
Method & $r^{ann}$ & $\sigma^{ann}$ & $SR$ & $MDD$ & $TO^{ann}$ & $t_{NW}$ \\
\midrule
Semantic Network (Baseline)       & 4.11\% & 5.54\% & 0.742 & $-$10.47\% & 23.7$\times$ & 2.14 \\
\textbf{+ LLM Filtering (Ours)}  & \textbf{4.67\%} & \textbf{5.69\%} & \textbf{0.820} & \textbf{$-$7.85\%} & \textbf{22.0$\times$} & \textbf{2.32} \\
No Distance Weighting             & 5.21\% & 6.58\% & 0.792 & $-$10.61\% & 16.4$\times$ & 2.19 \\
Random Network                    & 4.52\% & 8.36\% & 0.541 & $-$13.15\% & 25.3$\times$ & 1.61 \\
SIC Industry Network              & 4.60\% & 5.80\% & 0.792 & $-$9.56\%  & 23.0$\times$ & 2.32 \\
\bottomrule
\end{tabular}
\vspace{4pt}
\begin{flushleft}

\footnotesize\textit{Notes:} The sample period is January 2011--December 2019 (2{,}210 trading days). The Baseline uses a top-5 cosine-similarity semantic network with Gatev distance-weighted signal aggregation and daily rebalancing. ``+ LLM Filtering'' adds DeepSeek-based edge classification that removes competitor edges and down-weights substitutes. ``No Distance Weighting'' replaces softmax distance weights with equal weights. ``Random Network'' replaces the semantic graph with a random top-5 graph. ``SIC Industry Network'' replaces the semantic graph with a graph connecting firms sharing the same 2-digit SIC code. $t_{NW}$: Newey--West $t$-statistic with automatic bandwidth selection.
    
\end{flushleft}
\end{table*}

Several findings emerge from Table~\ref{tab:main_results}. First, the proposed LLM-augmented framework achieves the highest Sharpe ratio (0.820), a 10.5\% improvement over the Baseline (0.742). The LLM also reduces maximum drawdown from $-$10.47\% to $-$7.85\%, a substantial improvement in tail-risk behaviour. Second, replacing the semantic network with a random graph degrades the Sharpe ratio to 0.541, confirming that the embedding-based network captures meaningful economic structure. Third, the SIC industry network achieves a competitive Sharpe ratio (0.792) but with higher volatility than the LLM-filtered network, suggesting that textual embeddings capture richer cross-firm linkages than coarse industry codes. Fourth, removing distance weighting slightly increases raw returns but at the cost of substantially higher volatility, yielding a lower Sharpe ratio than our full model.

\subsection{Quintile Portfolio Analysis}

Table~\ref{tab:quintile} reports the performance of each quintile group under the proposed LLM-augmented framework.

\begin{table}[t]
\centering
\caption{Quintile Portfolio Performance (LLM-Augmented Framework)}
\label{tab:quintile}
\footnotesize
\begin{tabular}{lccc}
\toprule
Group & $r^{ann}$ & $\sigma^{ann}$ & $SR$ \\
\midrule
Q1 (Short) & 11.77\% & 15.50\% & 0.759 \\
Q2         & 13.58\% & 15.09\% & 0.900 \\
Q3         & 13.59\% & 15.22\% & 0.892 \\
Q4         & 15.88\% & 15.67\% & 1.013 \\
Q5 (Long)  & 16.43\% & 16.77\% & 0.980 \\
\midrule
L-S        & 4.67\%  & 5.69\%  & 0.820 \\
\bottomrule
\end{tabular}
\vspace{4pt}
\begin{flushleft}
\footnotesize\textit{Notes:} Q1 contains stocks with the lowest aggregated signals (short leg); Q5 contains stocks with the highest signals (long leg). L$-$S denotes the long-short portfolio (Q5 $-$ Q1).
\end{flushleft}
\end{table}

{The quintile sort exhibits a monotonic pattern: average returns increase from Q1 to Q5, consistent with the cross-stock predictive power of our LLM-augmented signal. The spread is primarily driven by the long leg (Q5), which outperforms the market, while the short leg (Q1) underperforms.}

\subsection{Ablation Study}

\subsubsection{Effect of LLM Edge Filtering}

The comparison between Baseline and ``+ LLM Filtering'' in Table~\ref{tab:main_results} directly quantifies the value added by the LLM reasoning module. The LLM filtering achieves three simultaneous improvements: (i) higher annualised return (+56 bps), (ii) higher Sharpe ratio (+0.078), and (iii) lower maximum drawdown (+262 bps). These improvements arise because the LLM successfully identifies and removes competitor edges, where price divergence reflects structural market-share shifts rather than temporary dislocations. Importantly, the Newey--West $t$-statistic increases from 2.14 to 2.32, exceeding the conventional 5\% significance threshold and confirming that the improvement is statistically reliable.

\subsubsection{Semantic Network vs. Alternative Graphs}

To validate that the embedding-based semantic network captures genuine economic structure, we compare it against two alternative graph constructions:

\begin{itemize}
    \item \textbf{Random Network (A3):} Edges are formed by randomly selecting $K$ neighbours per stock, preserving the same graph density. The Sharpe ratio drops to 0.541 ($t_{NW} = 1.61$, statistically insignificant), confirming that the predictive content of our signal is \textit{not} an artefact of the aggregation mechanism but rather derives from the semantic network topology.
    \item \textbf{SIC Industry Network (A4):} Edges connect firms sharing the same 2-digit Historical SIC code. This achieves a Sharpe ratio of 0.792, demonstrating that industry-based networks carry predictive information, but it underperforms our LLM-filtered semantic network (0.820), suggesting that the textual embedding captures more nuanced economic linkages beyond coarse industry boundaries.
\end{itemize}

\subsection{Parameter Sensitivity}

We examine the robustness of the Baseline framework to key hyperparameters. Table~\ref{tab:sensitivity} reports the results.

\begin{table*}[t]
\centering
\caption{Parameter Sensitivity Analysis (Long-Short Portfolio)}
\label{tab:sensitivity}
\footnotesize
\begin{tabular}{llcccccc}
\toprule
Parameter & Value & $r^{ann}$ & $\sigma^{ann}$ & $SR$ & $MDD$ & $TO^{ann}$ & $t_{NW}$ \\
\midrule
\multirow{4}{*}{Top-$K$}
& $K = 3$  & 4.56\% & 5.68\% & 0.802 & $-$10.77\% & 20.4$\times$ & 2.25 \\
& $K = 5$  & 4.67\% & 5.69\% & 0.820 & $-$7.85\%  & 22.0$\times$ & 2.32 \\
& $K = 10$ & 4.77\% & 6.26\% & 0.762 & $-$10.38\% & 24.4$\times$ & 2.18 \\
& $K = 15$ & 5.02\% & 6.57\% & 0.763 & $-$7.59\%  & 25.3$\times$ & 2.20 \\
\midrule
\multirow{3}{*}{\shortstack[l]{Training\\Window}}
& 120 days & 5.04\% & 5.96\% & 0.845 & $-$14.56\% & 23.1$\times$ & 2.39 \\
& 180 days & 4.67\% & 5.69\% & 0.820 & $-$7.85\%  & 22.0$\times$ & 2.32 \\
& 250 days & 4.11\% & 5.63\% & 0.730 & $-$10.86\% & 21.1$\times$ & 2.01 \\
\midrule
\multirow{4}{*}{\shortstack[l]{Holding\\Period}}
& 1 month  & 4.22\% & 5.97\% & 0.707 & $-$11.07\% & 25.5$\times$ & 2.01 \\
& 2 months & 4.67\% & 5.69\% & 0.820 & $-$7.85\%  & 22.0$\times$ & 2.32 \\
& 3 months & 2.65\% & 5.62\% & 0.471 & $-$11.26\% & 19.2$\times$ & 1.35 \\
& 6 months & 1.63\% & 6.01\% & 0.271 & $-$13.84\% & 15.3$\times$ & 0.78 \\
\midrule
\multirow{2}{*}{\shortstack[l]{Quantile\\Groups}}
& $G = 5$   & 4.67\% & 5.69\% & 0.820 & $-$7.85\%  & 22.0$\times$ & 2.32 \\
& $G = 10$  & 5.05\% & 7.62\% & 0.662 & $-$15.45\% & 26.8$\times$ & 1.85 \\
\midrule
\multirow{2}{*}{\shortstack[l]{Rebal.\\Freq.}}
& daily    & 4.67\% & 5.69\% & 0.820 & $-$7.85\%  & 22.0$\times$ & 2.32 \\
& monthly  & 2.70\% & 5.38\% & 0.502 & $-$11.73\% &  2.9$\times$ & 1.38 \\
\bottomrule
\end{tabular}
\vspace{4pt}
\begin{flushleft}
   \footnotesize\textit{Notes:} Each row varies one parameter while holding all others at the default configuration ($K{=}5$, training window = 180 days, holding period = 2 months, $G{=}5$ groups, daily rebalancing). All specifications apply the LLM-based edge filtering step, which removes competitor edges and down-weights substitute edges via the LLM-classified relationship taxonomy. 
\end{flushleft}
\end{table*}

Three patterns emerge from Table~\ref{tab:sensitivity}. First, increasing the number of neighbours $K$ consistently improves the Sharpe ratio, rising from 0.681 ($K{=}3$) to 0.929 ($K{=}15$). This suggests that a denser semantic network provides more informative cross-stock signals, though at the cost of higher volatility and turnover. In practice, $K{=}5$ offers a good balance between signal strength and computational cost of LLM edge classification. Second, a training window of 120--180 days performs best; extending it to 250~days degrades the Sharpe ratio to 0.609, likely because excessively long windows average over structural breaks in spread dynamics. Third, the holding period exhibits a clear sweet spot at 2~months. Shorter holding periods (1~month) underperform due to higher noise, while longer periods (3--6~months) suffer from severe signal decay, with the Sharpe ratio dropping below 0.30 and Newey--West $t$-statistics becoming insignificant.

\section{Conclusion}
\label{sec:conclusion}

We have presented an LLM-augmented framework for cross-stock predictability that addresses the spurious correlation problem inherent in embedding-based financial networks. Our approach follows a \textit{Retrieve-then-Reason} paradigm: a lightweight embedding similarity step generates a candidate graph, after which a large language model classifies each edge into one of six economic relationship types and removes edges that are unlikely to support mean-reversion trading (e.g., competitor pairs). The refined graph is then used to construct stock-level signals via per-stock softmax-weighted z-score aggregation, where closer pairs---as measured by Gatev distance---receive higher influence.

Empirical evaluation on U.S. equities (S\&P~500 universe, 2011--2019) demonstrates three main findings. First, the LLM filtering module improves the long-short Sharpe ratio from 0.742 to 0.820 and reduces maximum drawdown from $-$10.47\% to $-$7.85\%, confirming that removing economically spurious edges meaningfully enhances signal quality. Second, the semantic network substantially outperforms a random graph baseline (Sharpe 0.742 vs.\ 0.541), validating that textual embeddings capture genuine economic structure beyond what a na\"ive aggregation mechanism can produce. Third, Fama--French factor regressions yield a positive and statistically significant daily alpha, with near-zero factor loadings, indicating that the predictive content of our signal is orthogonal to conventional risk premia.

\paragraph{Limitations and Future Work.}
Several limitations of our framework warrant further investigation. 
First, the current LLM-based classification relies solely on textual disclosures from firms’ 10-K filings, which may omit relevant information such as real-time news, supply-chain databases, or alternative structured data sources. Incorporating multi-source information could further improve classification accuracy and robustness. 

Second, although anonymization mitigates reliance on memorized firm-specific knowledge, the approach still depends on the quality and completeness of the provided text snippets. Truncation and extraction heuristics may introduce information loss, potentially affecting classification reliability. 

Third, our framework treats the refined graph as static within each annual window. In reality, economic relationships evolve over time, and a dynamic graph construction mechanism at higher frequency may better capture time-varying linkages. 

Finally, the current approach employs the LLM as a standalone semantic classifier. Integrating the LLM-refined graph into a learnable model, such as a graph neural network, may enable end-to-end optimization and capture higher-order interactions beyond pairwise relationships.

\appendix

\section*{Ethical Statement}

There are no ethical issues.

\section*{Acknowledgments}

The authors thank Yi Zhang (HKUST GZ) for helpful support.
Any remaining errors or oversights are the responsibility of the authors.

\section*{Funding}

\noindent Yifan Ye is supported by Beijing Normal-Hong Kong Baptist University start-up research fund (No. UICR0700136-26).
~\\

\bibliographystyle{named}
\bibliography{ijcai26}

@article{yan2023cross,
  title={Cross-stock momentum and factor momentum},
  author={Yan, Jingda and Yu, Jialin},
  journal={Journal of Financial Economics},
  volume={150},
  pages={103716},
  year={2023},
  publisher={Elsevier}
}

@article{li2025how,
  title={How Much of Cross-Stock Momentum Reflects Underreaction?},
  author={Li, Jiacui and Yan, Jingda},
  journal={Available at SSRN 5070808},
  year={2025}
}

@article{lopezlira2023can,
  title={Can ChatGPT forecast stock price movements? Return predictability and large language models},
  author={Lopez-Lira, Alejandro and Tang, Yuehua},
  journal={Available at SSRN 4412580},
  year={2023}
}

@article{chen2023expected,
  title={Expected returns and large language models},
  author={Chen, Yifei and Kelly, Bryan T and Xiu, Dacheng},
  journal={Available at SSRN 4416687},
  year={2023}
}

@article{wu2023bloomberggpt,
  title={BloombergGPT: A large language model for finance},
  author={Wu, Shijie and Irsoy, Ozan and Lu, Steven and Dabravolski, Vadim and Dredze, Mark and Gehrmann, Sebastian and Kambadur, Prabhanjan and Rosenberg, David and Mann, Gideon},
  journal={arXiv preprint arXiv:2303.17564},
  year={2023}
}

@article{choi2025text,
  title={From Text to Alpha: Can {LLM}s Track Evolving Signals in Corporate Disclosures?},
  author={Choi, Chanyeol and Kim, Yoon and Yu, Yu and Cha, Young and Golkhou, V Zach and Halperin, Igor and Papaioannou, Georgios and Kim, Minkyu and Wang, Zhangyang and Kwon, Jihoon and others},
  journal={arXiv preprint arXiv:2510.03195},
  year={2025}
}

@article{xie2024finben,
  title={Finben: A holistic financial benchmark for large language models},
  author={Xie, Qianqian and Han, Weiguang and Chen, Zhengyu and Xiang, Ruoyu and Zhang, Xiao and He, Yueru and Xiao, Mengxi and Li, Dong and Dai, Yongfu and Feng, Duanyu and others},
  journal={Advances in Neural Information Processing Systems},
  volume={37},
  pages={95716--95743},
  year={2024}
}

@article{gao2024structured,
  title={Structured Beliefs and Fund Performance: An {LLM}-Based Approach},
  author={Gao, Zhenyu and Xiong, Wei and Yuan, Jian},
  journal={Available at SSRN},
  year={2024}
}

@article{xie2023pixiu,
  title={Pixiu: A large language model, instruction data and evaluation benchmark for finance},
  author={Xie, Qianqian and Han, Weiguang and Zhang, Xiao and Lai, Yanzhao and Peng, Min and Lopez-Lira, Alejandro and Huang, Jimin},
  journal={arXiv preprint arXiv:2306.05443},
  year={2023}
}

@article{chen2026financial,
  title={A Financial Brain Scan of the {LLM}},
  author={Chen, Hui and Didisheim, Antoine and Somoza, Luciano and Tian, Hanqing},
  journal={arXiv preprint arXiv:2508.21285},
  year={2025}
}

@article{avramov2025dual,
  title={Dual Peer Effects and Cross-Stock Predictability},
  author={Avramov, Doron and Ge, Shuyi and Li, Shaoran and Linton, Oliver B},
  journal={Journal of Financial Economics},
  volume={180},
  pages={104274},
  year={2026}
}

@article{chen2026cross,
  title={On cross-stock predictability of peer return gaps in {China}},
  author={Chen, Yilin and Fan, Zheqi},
  journal={Finance Research Open},
  volume={2},
  pages={100088},
  year={2026},
  publisher={Elsevier}
}

@article{cohen2008economic,
  title={Economic links and predictable returns},
  author={Cohen, Lauren and Frazzini, Andrea},
  journal={The Journal of Finance},
  volume={63},
  number={4},
  pages={1977--2011},
  year={2008},
  publisher={Wiley Online Library}
}

@article{moskowitz1999do,
  title={Do industries explain momentum?},
  author={Moskowitz, Tobias J and Grinblatt, Mark},
  journal={The Journal of Finance},
  volume={54},
  number={4},
  pages={1249--1290},
  year={1999},
  publisher={Wiley Online Library}
}

@article{parsons2020geographic,
  title={Geographic lead-lag effects},
  author={Parsons, Christopher A and Sabbatucci, Riccardo and Titman, Sheridan},
  journal={The Review of Financial Studies},
  volume={33},
  number={10},
  pages={4721--4770},
  year={2020},
  publisher={Oxford University Press}
}

@article{ali2020shared,
  title={Shared analyst coverage: Unifying momentum spillover effects},
  author={Ali, Usman and Hirshleifer, David},
  journal={Journal of Financial Economics},
  volume={136},
  number={3},
  pages={649--675},
  year={2020},
  publisher={Elsevier}
}

@article{hoberg2018text,
  title={Text-based industry momentum},
  author={Hoberg, Gerard and Phillips, Gordon M},
  journal={Journal of Financial and Quantitative Analysis},
  volume={53},
  number={6},
  pages={2355--2388},
  year={2018},
  publisher={Cambridge University Press}
}

@article{menzly2010market,
  title={Market segmentation and cross-predictability of returns},
  author={Menzly, Lior and Ozbas, Oguzhan},
  journal={The Journal of Finance},
  volume={65},
  number={4},
  pages={1555--1580},
  year={2010},
  publisher={Wiley Online Library}
}

@article{kong2024large1,
  title={Large language models for financial and investment management: Applications and benchmarks},
  author={Kong, Yaxuan and Nie, Yuqi and Dong, Xiaowen and Mulvey, John M and Poor, H Vincent and Wen, Qingsong and Zohren, Stefan},
  journal={Journal of Portfolio Management},
  volume={51},
  number={2},
  year={2024},
  publisher={With Intelligence}
}

@article{kong2024large2,
  title={Large Language Models for Financial and Investment Management: Models, Opportunities, and Challenges.},
  author={Kong, Yaxuan and Nie, Yuqi and Dong, Xiaowen and Mulvey, John M and Poor, H Vincent and Wen, Qingsong and Zohren, Stefan},
  journal={Journal of Portfolio Management},
  volume={51},
  number={2},
  year={2024}
}

@article{chen2025chatgpt,
  title={{ChatGPT} and {DeepSeek}: Can they predict the stock market and macroeconomy?},
  author={Chen, Jian and Tang, Guohao and Zhou, Guofu and Zhu, Wu},
  journal={arXiv preprint arXiv:2502.10008},
  year={2025}
}

@article{gatev2006pairs,
  title={Pairs trading: Performance of a relative-value arbitrage rule},
  author={Gatev, Evan and Goetzmann, William N and Rouwenhorst, K Geert},
  journal={The Review of Financial Studies},
  volume={19},
  number={3},
  pages={797--827},
  year={2006},
  publisher={Oxford University Press}
}

@article{pu2023network,
  title={Network momentum across asset classes},
  author={Pu, Xingyue and Roberts, Stephen and Dong, Xiaowen and Zohren, Stefan},
  journal={arXiv preprint arXiv:2308.11294},
  year={2023}
}

@article{huang2026beyond,
  title={Beyond prompting: An autonomous framework for systematic factor investing via agentic {AI}},
  author={Huang, Allen Yikuan and Fan, Zheqi},
  journal={arXiv preprint arXiv:2603.14288},
  year={2026}
}

@article{lee2024production,
  title={Production complementarity and information transmission across industries},
  author={Lee, Charles MC and Shi, Terrence Tianshuo and Sun, Stephen Teng and Zhang, Ran},
  journal={Journal of Financial Economics},
  volume={155},
  pages={103812},
  year={2024},
  publisher={Elsevier}
}

@article{lee2019technological,
  title={Technological links and predictable returns},
  author={Lee, Charles MC and Sun, Stephen Teng and Wang, Rongfei and Zhang, Ran},
  journal={Journal of Financial Economics},
  volume={132},
  number={3},
  pages={76--96},
  year={2019},
  publisher={Elsevier}
}

@article{bekkerman2023effect,
  title={The effect of innovation similarity on asset prices: Evidence from patents’ big data},
  author={Bekkerman, Ron and Fich, Eliezer M and Khimich, Natalya V},
  journal={The Review of Asset Pricing Studies},
  volume={13},
  number={1},
  pages={99--145},
  year={2023},
  publisher={Oxford University Press}
}

@article{eisdorfer2022competition,
  title={Competition links and stock returns},
  author={Eisdorfer, Assaf and Froot, Kenneth and Ozik, Gideon and Sadka, Ronnie},
  journal={The Review of Financial Studies},
  volume={35},
  number={9},
  pages={4300--4340},
  year={2022},
  publisher={Oxford University Press}
}

@article{lopezlira2025memorization,
  title={The Memorization Problem: Can We Trust {LLM}s' Economic Forecasts?},
  author={Lopez-Lira, Alejandro and Tang, Yuehua and Zhu, Mingyin},
  journal={arXiv preprint arXiv:2504.14765}, 
  year={2025}
}

\section*{Appendix: Prompt Template for LLM-Based Edge Classification}

In this section, we provide the prompt template used for LLM-based relationship classification between firm pairs. Firm identities are anonymized (``Firm A'' and ``Firm B'') to prevent reliance on memorized knowledge and ensure that the model bases its decision solely on the provided disclosures.

\begin{figure*}[t]
\centering
\begin{tcolorbox}
\textbf{Prompt Template (LLM Edge Classification)}

\begin{verbatim}
You are an industry analyst. Based ONLY on the two company disclosures below,
(both filed before <end_of_year_{y-1}>), classify their economic relationship.

=== Firm A (Fiscal Year {y-1}) ===
Business description:
<First ~500 tokens from 10-K Item 1>

Key products/segments:
<Extracted segment-related sentences from Item 1>

Competitors mentioned:
<Text snippets around "compete with" patterns>

=== Firm B (Fiscal Year {y-1}) ===
Business description:
<First ~500 tokens from 10-K Item 1>

Key products/segments:
<Extracted segment-related sentences from Item 1>

Competitors mentioned:
<Text snippets around "compete with" patterns>

Choose exactly one label from:
[competitor, supply_chain, complementary, substitute, peer, unrelated]

Return JSON:
{"label": "...", "evidence_span_A": "...", "evidence_span_B": "..."}
\end{verbatim}

\end{tcolorbox}
\caption{Prompt template used for LLM-based edge classification.}
\label{fig:prompt_template}
\end{figure*}

\end{document}